\documentclass[longbibliography,reprint,aps,prx,amsmath,amssymb,superscriptaddress]{revtex4-2} 
\usepackage[utf8]{inputenc}
\usepackage[T1]{fontenc}
\usepackage{multirow}
\usepackage{subcaption}
\usepackage{tabularx}
\usepackage{graphicx,caption} 
\usepackage{amsfonts,amsmath,amssymb}
\usepackage{comment}
\usepackage{lineno}
\usepackage{hyperref}
\usepackage[explicit]{titlesec}

\newcommand{\be}{\begin{equation}}
\newcommand{\ee}{\end{equation}} 
\newcommand{\bea}{\begin{eqnarray}}
\newcommand{\eea}{\end{eqnarray}}

\renewcommand{\bf}[1]{\textbf{#1}} 

\newcommand{\f}[2]{\frac{#1}{#2}}

\newcommand{\ccup}[1]{\left\{#1\right\}}

\newcommand{\bup}[1]{\left(#1\right)}
\newcommand{\rup}[1]{\left[#1\right]}

\newcommand{\Exp}{\mathbb{E}}

\renewcommand{\ref}[1]{[\ref{#1}]}

\usepackage{algorithm,algorithmic}
\usepackage[algoruled,algo2e]{algorithm2e}
\usepackage{sidecap}
\usepackage{boldline}
\usepackage{setspace}
\usepackage{listings}
\usepackage{fancyvrb}
\usepackage{soul}

\usepackage{mathtools}

\usepackage[table]{}
\usepackage{array}

\usepackage[utf8]{inputenc}

\usepackage{booktabs}
\usepackage{adjustbox}

\usepackage{changepage}

\usepackage[usenames,dvipsnames]{xcolor}
\definecolor{shadecolor}{gray}{0.9}
\usepackage{tabu}

\usepackage[normalem]{ulem} 

\newcommand{\mtrep}{\mbox{{\small CRep}}}
\newcommand{\dmtrep}{\mbox{{\small DynCRep}}}
\newcommand{\dmtrepo}{\mbox{{\small DynCRep$_{0}$}}}

\newcommand{\sbm}{\mbox{{\small SBM}}}
\newcommand{\dsbm}{\mbox{{\small DSBM}}}
\newcommand{\wdyn}{\mbox{{\small w-DYN}}}
\newcommand{\eud}{\mbox{{\small EU-daily}}}
\newcommand{\eum}{\mbox{{\small EU-monthly}}}
\newcommand{\wsta}{\mbox{{\small w-STATIC}}}
\graphicspath{{figures/}} 
\usepackage{cleveref}
\crefname{equation}{Eq.}{Eqs.}
\crefname{section}{Sec.}{Secs.}
\crefname{figure}{Fig.}{Figs.}

\begin{document}

\title{Reciprocity, community detection, and  link prediction  in dynamic networks}

\author{Hadiseh Safdari}
	\email{hadiseh.safdari@tuebingen.mpg.de} 
	\affiliation{Max Planck Institute for Intelligent Systems, Cyber Valley, Tuebingen 72076, Germany}

\author{Martina Contisciani}
	\email{martina.contisciani@tuebingen.mpg.de} 
	\affiliation{Max Planck Institute for Intelligent Systems, Cyber Valley, Tuebingen 72076, Germany}

\author{Caterina De Bacco}
	\email{caterina.debacco@tuebingen.mpg.de} 
	\affiliation{Max Planck Institute for Intelligent Systems, Cyber Valley, Tuebingen 72076, Germany}

\begin{abstract}
Many complex systems change their structure over time, in these cases dynamic networks can provide a richer representation of such phenomena. As a consequence, many inference methods have been generalized to the dynamic case with the aim to model dynamic interactions. Particular interest has been devoted to extend the stochastic block model and its variant, to capture community structure as the network changes in time. While these models assume that edge formation depends only on the community memberships, recent work for static networks show the importance to include additional parameters capturing structural properties, as reciprocity for instance. Remarkably, these models are capable of generating more realistic network representations than those that only consider community membership. To this aim, we present a probabilistic generative model with hidden variables that integrates reciprocity and communities as structural information of networks that evolve in time. The model assumes a fundamental order in observing reciprocal data, that is an edge is observed, conditional on its reciprocated edge in the past. We deploy a Markovian approach to construct the network's transition matrix between time steps and parameters' inference is performed with an Expectation-Maximization algorithm that leads to high computational efficiency because it exploits the sparsity of the dataset. We test the performance of the model on synthetic dynamical networks, as well as on real networks of citations and email datasets.  We show that our model captures the reciprocity of real networks better than standard models with only community structure, while performing well at link prediction tasks. 
\end{abstract}

\maketitle 
\section{Introduction}
Many real networks are dynamical, i.e., the pattern of interactions between their nodes vary over time,  e.g., network of exchanged emails in a company. The abundance of such datasets and the development  of optimal numerical methods have led to a growing number of studies in this field \cite{Vespignani:2012aa, ISI:000365733100002, 10.1145/3097983.3098145,Mandjes2019}.   In addition, interactions between nodes can be reciprocated,  e.g., the people whom one retweets and the number of times she retweets them vary over time; so do the papers that researchers cite in their manuscripts and papers that cite one's scientific output. This latter issue has received little attention in  previous studies.

Among the main approaches to study these systems, latent variable models assume that the existence of an edge between any pair of nodes is independent of other nodes, and is conditional on some latent variables which incorporate the hidden structure of the network. 
 These techniques mainly focus on community membership as the main relevant latent variable, e.g., in the case of citations, the people who cite each other's works, inadvertently  form a community. The stochastic block model (\sbm\text{}) \cite{holland1983stochastic, Yuchung1987,10.2307/2670253}  and its variants provide flexible network generative models \cite{KarrerNewmanPRE2011,de2017community}. In this framework, nodes are initially partitioned into communities, then edges are created between nodes, based on their community  membership. There are several variants of dynamical equivalents of stochastic block model (\dsbm\text{}) \cite{yang2011detecting, matias2015statistical,zhang2017random,xu2014dynamic,han2014consistent} which capture transition of community membership over time, reflecting the evolution of edge formation. \citet{peixoto2017modelling}, and  \citet{matias2018} develop a non-parametric temporal stochastic block model. \citet{gauvin2014detecting} consider non-negative tensor factorization, where communities are static but the affinity matrix changes over time. \citet{bovet2021flow} use flow of random walkers co-evolving in the dynamic network to define communities.  Various methods have been used to address whether the community membership or connectivity parameters could change over time, see \cite{rossetti2018community} for a review. For instance, one could assume that communities are fixed in time but the connectivity parameters across groups changes, as in \cite{gauvin2014detecting,matias2015statistical}, or  that communities change in time \cite{yang2011detecting,ghasemian2016detectability,mucha2010community,herlau2013modeling}.

In Zhang et al. \cite{zhang2017random}, the authors  extend some of  the popular methods of modeling network structure, e.g., \sbm \text{},  to represent  dynamic networks. The main idea behind their Markovian approach is to find transition rates of appearance and disappearance of edges over time. Based on these rates, they were able to calculate  the average probability of edges over all time steps, hence, they estimate a steady state probability distribution  for  each network model, depending on its structural  parameters. Although  the approach followed in \cite{zhang2017random} is efficient and analytically grounded, it was developed for  models that  incorporate communities  as the only latent variable. 

Nevertheless, in directed real-world  networks, community  membership may not be the only factor influencing network structure. Reciprocity, i.e., the tendency of a pair of nodes to form edges on both directions, has  been subject of  many studies \cite{holland1981exponential, PhysRevLett.93.268701, safdari2020generative} as a crucial factor to determine the structure of networks, in particular in social networks.
\citet{Bartolucci:2018aa} assume local conditional independence between pairs of edges,  i.e., dyads,  and extend the \sbm \text{} to account for the reciprocal patterns in directed dynamical networks. 
Furthermore, they established various  specifications of the proposed model  corresponding to different reciprocal assumptions. 

Recently, a generative model (\mtrep\text{}) has been  introduced that, in addition to community membership, includes reciprocity as latent variable that dictates formation of edges between the nodes  \citep{safdari2020generative}. In other words, the appearance of a directed  edge from node $i$ to $j$ not only depends on the community that the nodes belong to, but also is affected by the existence of the edge from $j$ to $i$.  In the case of citation network, it is more likely for an author to cite those other who already cited her, implying overlapping research areas.

In this work, following the approach in \cite{zhang2017random}, we extend \mtrep\text{} and propose a continuous-time Markov process model for dynamic networks (\dmtrep \text{}). Observing the system at discrete points in time, at each time step the transition rates of  appearance and disappearance  of a directed edge between two nodes  depends on the current community membership of the nodes, as well as on  the existence of a reciprocated edge between them. 

We validate  the applicability of the proposed model and  its inference approach by performing experiments on real and synthetic networks for community detection and link prediction.
We apply the model  to synthetic  datasets and observe that  \dmtrep \text{} shows a reasonable performance in terms of link prediction.  Moreover, we test the model performance on real-world datasets in the domain of social and online communication to reproduce reciprocity, with promising results.  

\section{Model}
In our model, the temporally evolving network is captured in snapshots taken at fixed intervals, from $t=0$ to $T+1$.  
$A(t)$ represents the dynamic adjacency matrix of the network, where  a non-zero value of $A_{ij}(t)$  represents a weighted edge from $i$ to node $j$ at time $t$, and $A_{ij}(t)=0$  denotes no interaction. 
We assume that the total number of nodes is fixed over time, i.e., new nodes do not enter into the network, and nodes do not leave it; instead,  existing edges can appear and disappear. We focus on directed, and weighted networks.

A matrix $w(t)$ of dimensions $K \times K$  determines the evolving  structure of the $K$ communities over time and we refer to  $w(t)$  as the  affinity matrix.  Different assumptions about $w(t)$ result in communities with different structures.  For instance, in the case of diagonal entries being  greater than off-diagonal ones, communities are assortative -- that is, individuals are more inclined towards  intra-community interactions  than inter-community interactions. The $K$-dimensional vectors $u_{i}(t)$ and $v_{i}(t)$ denote the out-going and  in-coming  communities at time step $t$, respectively.

Here, we keep  the community membership constant over time; hence, we drop the notion of time dependency. We develop the model in two different varieties: 1) the affinity matrix varies over time (\wdyn), i.e., the connectivity pattern between communities changes over time, for instance, a group of nodes which form a community at time step $t$ could be peripheral nodes at another time step \cite{matias2015statistical}, and 2) the affinity matrix also remains static (\wsta \text{}). 

Following the continuous-time Markov process approach in \cite{zhang2017random}, we assume that networks evolve on the real-valued times; hence, the appearance and disappearance of the edges are continuous parameters. However, we observe the network at discrete time steps. At each time step,  a Poisson distribution governs the existence of edges between nodes  such  that an  edge between two nodes is formed at a rate  $\hat{\lambda}_{ij}(t)$. This rate depends on both the community that nodes belong to, and the existence of the reciprocated tie at the previous time step:
   \bea
  \hat{\lambda}_{ij}(t)&&=\lambda_{ij}(t)+\eta \, A_{ji}(t-1)   \nonumber \\
    && \equiv \sum_{k,q}u_{ik}v_{jq}\,w_{kq}(t)+ \eta \, A_{ji}(t-1) \quad,
    \label{eq:rappear}
   \eea
where  $\eta$ as a hyperparameter regulates the reciprocity effects, similarly as in \citep{safdari2020generative}. The difference between Eq. \ref{eq:rappear}  and  the edge probability in  \citep{safdari2020generative} is that the dependency on the reciprocated tie is on the previous time step, while standard CRep considers only the same time $t$, being an approach valid for static networks. Furthermore, an edge could disappear with rate  $\mu$.

\subsection{Dynamic \mtrep \text{}}
\label{sec:dyncrep}
The aim of this study is to infer the  latent parameters of the  model, namely, $\Theta \equiv \{ u, v, w, \eta, \mu \}$, given the adjacency matrix observed at each time step. To this end, we perform this inference task by maximizing the log-likelihood. Given $\Theta$, all the  pairs of nodes are conditionally independent; as a result, the joint-probability of the node-pairs could be approximated by a factorized form. Here, we develop a Markov process, according to which, at every time step, the probability of edges depends only on the previous time step: 
\begin{small}
\bea
P(\{A(t)\}| \Theta )  &&= P\left(\{A(t)\}| \{A(t-1)\} ,\Theta \right)     \nonumber \\ \nonumber \\
&& =\prod_{i,j} \Bigg\{  P \left(A_{ij}(0)|A_{ji}(0),\Theta \right)   \Bigg. \nonumber \\
&& \Bigg.  \times \prod _{t=1}^T \{P(A_{ij}(t)|A_{ij}(t-1),A_{ji}(t-1),\Theta )\} \Bigg\} \quad .   \nonumber \\
\eea
\end{small} 
We further assume that at the initial time step the probability $A_{ij}(0)$ of an edge between two nodes follows a Poisson distribution with mean $ \hat{\lambda}_{ij}=\lambda_{ij}(0)$, i.e., there is  no  reciprocated edge in the past:
 
\bea\label{eqn:A0}
P (A_{ij}(0)|A_{ji}(0),\Theta )=\frac{e^ {\lambda_{ij}(0)}  \lambda_{ij}(0) ^{A_{ij}(0)}}{A_{ij}(0)!} \quad.
\eea

At each time-step, edges appear with rate $\hat{\lambda}_{ij}(t)$, and disappear with rate $\mu$. We follow an approach similar to that of Zhang et al. \cite{zhang2017random} and  calculate the probability of the existence of edges by solving a master equation. Defining $p_{ij}^k(t)$ as the probability of having $k$ edges, i.e., an edge with the weight equal to $k$, between nodes $i,j$ at time $t$, this quantity satisfies the following  master equation:
 
\bea\label{eqn:mst1}
&& \f{dp_{ij}^{k}(t)}{dt}   = \nonumber \\
&& \hat{\lambda}_{ij}(t) \, p_{ij}^{k-1}(t) + (k+1)  \mu p_{ij}^{k+1}(t)- \left(\hat{\lambda}_{ij}(t)+k \mu \right)p_{ij}^{k}(t) \quad . \nonumber \\
\eea
To solve this equation, we use a generating function approach \cite{gardiner2004handbook},  by defining $g(z,t)=\sum_{k=0}^{\infty} p^{k}(t) z^{k}$. The solution for the generating function,
\bea  
\label{eqn:gzt}
g(z,t) =  f\left[(z-1) e^{-\mu t}\right]  e^{\frac{(z-1) \hat{\lambda}_{ij}(t)}{\mu } } \quad, \qquad
\eea
could be expanded in terms of $z$ to give us $p_{ij}^t$ (more details in Sec. \ref{appendix:master_eq}). There are  four possible transitions from time $t-1$ to $t$: 1) there is no edge neither at time $t-1$, nor at $t$; 2)  the appearance of an edge from non-edge, 3) disappearance of an existing edge, and 4) an existing edge remains; with the following probabilities, respectively, 
\bea
&& p_{0\rightarrow 0} = e^{-\beta (\lambda_{ij}(t)+\eta \,A_{ji}(t))} \nonumber  \\
&& p_{0\rightarrow 1} =    \beta  (\lambda_{ij}(t)+\eta \,A_{ji}(t)) e^{-\beta   (\lambda_{ij}(t)+\eta \,A_{ji}(t))}\nonumber  \\
&& p_{1\rightarrow 0} =   \beta e^{-\beta (\lambda_{ij}(t)+\eta \,A_{ji}(t))} \nonumber  \\
&& p_{1\rightarrow 1} =    (1-\beta)  e^{-\beta  (\lambda_{ij}(t)+\eta \,A_{ji}(t))},
\eea
where $\beta=1- e^{-\mu}$.
This  leads to the following time-dependent, log-likelihood:
\bea \label{eq:Lconv0}
L(T, \Theta)&& =\log[P(\{A(t)\}|\{A(t-1)\},\Theta )]  \nonumber \\   \nonumber \\ 
&& =\sum _{i,j}  \left\{\log \left[e^{-\lambda_{ij}(t)} \lambda_{ij}(t)^{A_{ij}(0)}\right] \right.\nonumber \\ 
&& \left. +\sum _{t=1}^T \log \left[  e^{-\beta  \left(\lambda_{ij}(t)+\eta \, A_{ji}(t)\right)}  \right. \right. \nonumber \\ \nonumber \\ 
&& \left. \left.  \times  \left[\beta  \left( \lambda_{ij}(t)+\eta \, A_{ji}(t)\right)\right]{}^{\left(1-A_{ij}(t-1)\right) A_{ij}(t)} \right. \right. \nonumber \\ \nonumber \\ 
&& \left. \left.  \times  \beta ^{A_{ij}(t-1) \left(1-A_{ij}(t)\right)}  \times(1-\beta )^{A_{ij}(t-1) A_{ij}(t)}     \right]  \right\} \, . \qquad
\eea
\
\setlength{\textfloatsep}{5pt}
\begin{algorithm}[H]
\SetKwInOut{Input}{Input}
	\setstretch{0.7}
	\Input{network $A(t)=\{A_{ij}(t)\}_{i,j=1}^{N}$, \\number of communities $K$.}
  	\BlankLine
	\KwOut{membership $u=\rup{u_{ik}},\, v=\rup{v_{ik}}$; network affinity matrix $w(t)=\rup{w_{kq}(t)}$; reciprocity parameter $\eta$; edge disappearance rate $\beta(t)$.}
	\BlankLine
	 Initialize $u,v,w(t),\eta, \beta(t)$ at random. 
	 \BlankLine
	 Repeat until $\mathcal{L}$ converges:
	 \BlankLine
	\quad 1. Calculate $\rho_1(t)$ and $\phi(t)$ (E-step): 
	\bea
	 && \rho^{(1)}_{ij}(t)= \frac{\lambda_{ij}(t)}{\lambda_{ij}(t)+\eta \,A_{ji}(t)} \;,\quad\rho^{(2)}_{ij}(t)= \frac{\eta\,A_{ji}(t)}{\lambda_{ij}(t)+\eta\,A_{ji}(t)} \;,\quad \nonumber \\ &&\phi_{ijkq}(t) =\frac{u_{ik}v_{jq}w_{kq}(t) }{\sum _{k,q}  u_{ik} v_{jq} w_{kq}(t)} \;.\quad  
	\nonumber
	\eea
	 \quad 2. Update parameters $\Theta$ (M-step):  
	\BlankLine
	\quad \quad \quad 
		i) for each node $i$ and community $k$ update memberships:
		\bea
		\quad  u_{ik}= \frac{a-1 + \sum _{j,q,t}\,\rho^{(1)}_{ij}(t) \, \phi _{ijkq}(t)\, \hat{A}_{ij}(t) }{b + \sum _{j,q}   v _{jq}\, \sum_{t=0}^T  \hat{\beta}(t) \,w_{kq}(t) } \nonumber\\
\quad  v_{ik} =\frac{a-1 + \sum _{j,q,t}\, \rho^{(1)}_{ij}(t) \, \phi _{jiqk}(t)\, \hat{A}_{ij}(t) }{b + \sum _{j,q}   u_{jq} \, \sum_{t=0}^T  \hat{\beta}(t) \,w_{kq}(t)} \nonumber
		\eea
	\quad \quad \quad
	ii) for each pair $(k,q)$ update affinity matrix:
		\be
		\quad w_{kq}(t)=\frac{\sum _{i,j} \rho^{(1)}_{ij}(t)  \phi _{ijkq}(t) \hat{A}_{ij}(t)}{\sum _{i,j} u_{ik}\,v_{jq} \hat{\beta}(t ) } \nonumber
		 \ee
	\quad \quad \quad
		iii) update reciprocity parameter:
		\be
		\quad \eta =  \frac{\sum _{i,j,t}\rho^{(2)}_{ij}(t)  \hat{A}_{ij}(t)}{\sum _{i,j,t=1}  \hat{\beta}(t) \, A_{ji}(t-1)}\nonumber
		 \ee
	\caption{\dmtrep \text{} (\wdyn):  EM algorithm.}
	\label{alg:EM}
\end{algorithm}

We add parameters' regularization by assuming Gamma-distributed priors for the membership vectors:
\be
P(u_{ik}; a,b) \propto u_{ik}^{a-1}e^{-b u_{ik}} \quad,
\ee
where $a\geq 1$, to  ensure the maximization of the log-likelihood (the second derivative must be negative), similarly for the $v_{ik}$. This adds new terms to the log-likelihood:
\bea
\mathcal{L}(T, \Theta) &=& L(T, \Theta) + (a-1) \sum_{i,k}\log u_{ik} -b \sum_{ik}u_{ik} \nonumber\\
&&+ (a-1) \sum_{i,k}\log v_{ik} -b \sum_{ik}v_{ik}\quad. \label{eq:Lconv}
\eea
In the experiments below we set the values of the hyper-priors to enforce sparsity, i.e., $a=1.5$, $b=10$. \\
Maximizing $\mathcal{L}(T, \Theta)$ requires taking the derivative of Eq. (\ref{eq:Lconv}) w.r.t.  each parameter individually and setting them to zero. Because the summations in the logarithm  render the calculations difficult, we employ  a variational approximation using Jensen's inequality. 
Inference is then performed using the Expectation-Maximization algorithm (EM); details are provided in \cref{appendix:EM}.

Hitherto, we have included all the dependencies on the reciprocated edge $A_{ji}(t-1)$  by considering the previous time step $t-1$. However, the model still applies if we incorporate the reciprocated edge at the same time step, i.e.,  considering $A_{ji}(t)$.  This choice may depend on the application  itself based on the expectations and insight of the practitioner from the reciprocity effects. Alternatively, one can choose between these two options with model-selection criteria. In our experiments on real data we deployed them both, and presented the version that performs best in cross-validation tasks (\cref{appendixsec:lp}).

We continue with two specifications of the model with different assumptions on the temporal evolution of  the affinity matrix. In the first approach, \wdyn \text{}, the affinity matrix is treated as  a time-dependent variable; while the community membership vectors, $u_i, v_i$,  are  kept static over time. Notice that a similar scenario could be obtained by fixing $w$ and changing $u_i, v_i$  in time \cite{matias2015statistical}, our model can be easily adapted to accommodate this alternative interpretation. Our model assumes fixed number of communities $K$. As we consider a mixed-membership model, we have the flexibility of allowing nodes to belong to various communities and with various intensities, thus allowing to capture the likelihood of the data well by effectively changing how an entry $u_{ik}$ or $v_{ik}$ impacts the magnitude of $\lambda_{ij}(t)$ via $w(t)$ in the \wdyn \text{} scenario, while keeping $K$ constant.

In the second scenario, \wsta, the affinity matrix is kept static as well. The purpose of considering these scenarios is to make the model flexible in dealing with various community structures (see \cref{APXsec:static,APXsec:w-temp,APXsec:dyn} for more details on each scenarios). Notice that in the case of \wsta, although all the latent variables are fixed in time, the network can still evolve, as edges appear and disappear based on the parameters $\beta$ and $\mu$. This is also the case for the Markov model (without reciprocity)  in  \cite{zhang2017random}.

For instance, the EM algorithm for \wsta \text{} yields: 
\bea
&&u_{ik}=  \frac{a-1 + \sum _{j,q,t} \, \rho^{(1)}_{ij}(t)\, \phi _{ijkq}\, \hat{A}_{ij}(t) }{b + \sum _{j,q}   v _{jq} \,w_{kq}\, \bup{1+ \beta\,T}} \label{eq:LdifuStatic} \\
&&v_{jq} =\frac{a-1 + \sum _{i,k,t}\, \rho^{(1)}_{ij}(t) \, \phi _{ijkq}\, \hat{A}_{ij}(t) }{b + \sum _{i,k} \,   u_{ik}     \,w_{kq} \, \bup{1+ \beta\,T}} \label{eq:LdifvStatic}\\
&&w_{kq}=\frac{\sum _{i,j,t}\rho^{(1)}_{ij}(t)  \phi _{ijkq} \hat{A}_{ij}(t)}{\sum _{i,j}\,  u_{ik}\,v_{jq} \, \bup{1+ \beta\,T}} \label{eq:LdifwStatic} \\  \nonumber   \\
&&\eta=  \frac{\sum _{i,j,t}\rho^{(2)}_{ij}(t) \hat{A}_{ij}(t)}{\sum _{i,j}\,\sum _{t=1}^{T}\, \beta \, A_{ji}(t-1)}, \label{eq:LdifeStatic} \\ \nonumber    
\eea 
where we defined $\hat{A}_{ij}(t) =   A_{ij}(t) (1-A_{ij}(t-1))$ if $t>0$, in which $\hat{A}_{ij}(0)=A_{ij}(0)$ and we have the variational distributions
\bea 
&& \rho^{(1)}_{ij}(t)= \frac{\lambda_{ij}}{\lambda_{ij}+\eta \,A_{ji}(t-1)} \quad   \label{eqn:rho1static}  \\
&& \rho^{(2)}_{ij}(t)= \frac{\eta\,A_{ji}(t-1)}{\lambda_{ij}+\eta\,A_{ji}(t-1)} \quad  \label{eqn:rho2static}\\
&&   \phi_{ijkq} =\frac{u_{ik}v_{jq}w_{kq} }{\sum _{k,q}  u_{ik} v_{jq} w_{kq}} \quad. \label{eqn:phistatic}
\eea

The parameter $\beta$ has no closed-form update: 
\begin{small}
\begin{align}
-\beta \rup{T\sum_{i,j} \lambda_{ij}  + \sum_{i,j,t=1}^{t=T}\bup{\eta A_{ji}(t-1)}+\f{1}{1-\beta} A_{ij}(t-1) A_{ij}(t)}  \nonumber  \\
+ \sum_{i,j,t=1}^{t=T}\rup{\hat{A}(t)+A_{ij}(t-1)(1-A_{ij}(t))}=0 \,,\quad
\end{align}
\end{small}but this equation can be solved numerically using root-finding methods. 
The algorithm proceeds by randomly initializing the parameters $u,v,w,\eta,\beta$; then we estimate  the variational distributions $\rho^{(1)},\rho^{(2)}$, and $\phi$, using Eq.~(\ref{eqn:rho1static}) to (\ref{eqn:phistatic}) (E-step), while keeping  the parameters fixed. In the next step (M-step),  we update the parameters,  while keeping $\rho^{(1)},\rho^{(2)}$ and $\phi$ fixed. This procedure is repeated until the convergence of the likelihood in Eq.~(\ref{eq:Lconv}). An overview of the algorithm is described in \Cref{alg:EM}. 

\begin{figure*}[htb]
	\includegraphics[width=1\linewidth]{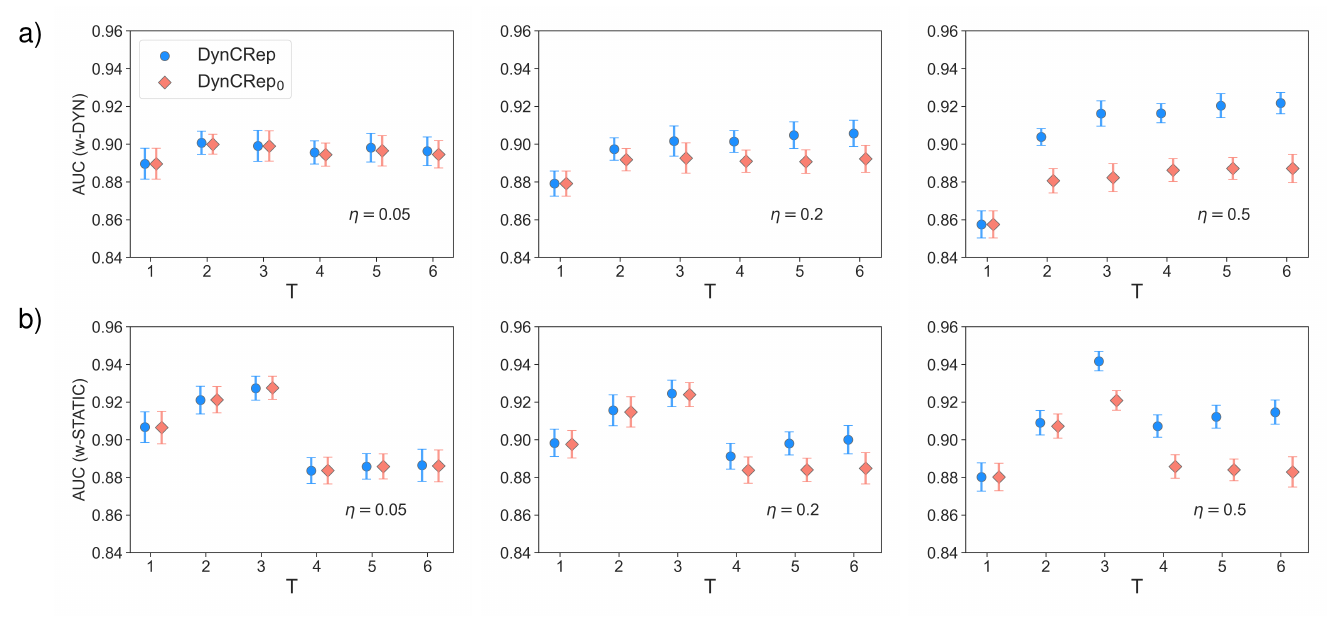}
	\caption{\textbf{Predicting future evolution. }  We report the AUC values on held-out experiments where we train the model on $A(0),\dots,A(T-1)$ and predict the network $A(T)$. Higher values means better prediction. Networks are generated as explained in \cref{sec:app}, with  $N=500$, average degree $\langle k \rangle=5$, $\beta=0.2$, $K=3$. The three plots are results for $\eta \in \ccup{0.05,0.2,0.5}$. The Markers and the error bars are the means and standard deviations over 20 network samples, respectively. a) \wdyn; b) \wsta.} 
	\label{fig:AUCsynth}
\end{figure*}

\section{Applications}\label{sec:app}
\subsection{Synthetic networks: AUC}
\label{sec:aucsynth}
Having explained the nots and bolts of our model, we now turn to its application on dynamic network data.
We start by considering synthetic networks generated by \cref{sec:dyncrep} with known community structure and reciprocity.
We assess the ability of the model in predicting the network at future time steps using past observations. 
We look in particular at the impact of reciprocity in determining edges, by generating networks with varying $\eta \in \ccup{0.05,0.2,0.5}$, while keeping the other parameters fixed. \\
For the tests reported here we use $N=500$, initial average degree $\langle k \rangle=5$, and $\beta=0.2$. We generate $K=3$ hard communities of equal size with assortative structure. Having fixed the parameters, we generate 20 samples of networks for each of the three values of $\eta$. For each network we generate an initial state followed by up to $T=6$ further snapshots. The initial state is generated using only the community structure (no reciprocity) using Eq.~(\ref{eqn:A0}). The successive snapshots are generated according to the instructions of \cref{sec:dyncrep}. In this study,  to test the ability of our model in capturing the dynamical features, we generate the first three time snapshots ($T=1,2,3$)  with an assortative community structure and the rest of the snapshots ($T=4,5,6$) with a disassortative community structure.\\

For each time step $t\in [1,T]$, we hide the individual snapshot $A(t)$ and fit the data using the previous snapshots $A(0),\dots,A(t-1)$. 
We test whether a model that accounts for reciprocity is able to successfully predict the network's evolution. Success is measured using the area under the curve (AUC), i.e., the probability that a randomly selected edge has higher expected value than a randomly selected non-existing edge. A value of $1$ means perfect reconstruction, while $0.5$ is pure random chance. The expected value of an edge is computed using:
\bea
\Exp\rup{ A_{ij}(t)} &=& \begin{cases} \f{p_{0\rightarrow 1}}{p_{0\rightarrow 1}+p_{0\rightarrow 0}} &\text{if} \,\, A_{ij}(t-1)=0\\
						 \f{p_{1\rightarrow 1}}{p_{1\rightarrow 1}+p_{1\rightarrow 0}} &\text{if} \,\,  A_{ij}(t-1)=1
						 \end{cases}	\\
						 &=& \begin{cases} \beta(t)(\lambda_{ij}(t)+\eta A_{ji}(t-1)) &\text{if} \,\, A_{ij}(t-1)=0\\
						 1-\beta(t) &\text{if} \,\,  A_{ij}(t-1)=1 
						 \end{cases}	\nonumber \,.
\eea 
Notice that while the expected value at time $t$ uses explicitly only the network at the previous time step, all the parameters are inferred using the \textit{whole} network history, i.e., the model is trained with  $\ccup{A(0),\dots, A(t-1)}$.  
We compare with a model that does not account for reciprocity, i.e., our model with $\eta=0$ (\dmtrepo) \cite{safdari2020generative}.\\

Figure \ref{fig:AUCsynth} shows the results of these tests. As we can see, the ability to predict future edges is greater for a model that accounts for
reciprocity, and the performance gap increases for higher values of $\eta$. This gap is partially offset by increasing the number of snapshots, as
both the models have access to more information to make their estimates.  
Remarkably, \dmtrep \text{} has stronger performance also in the low-reciprocity regime, $\eta=0.05$. This cannot be clearly seen by looking at \cref{fig:AUCsynth}, as the mean AUC of the two models are within the error bars due  to random fluctuations of the network structure across samples.  Instead, the stronger performance of \dmtrep \text{} in the low-reciprocity regime is revealed by looking at the percentage of samples where \dmtrep \text{} has higher AUC than \dmtrep0, on a trial-by-trial case (see \Cref{tab:synth_edge} for details). While \wsta \text{}, the static version of the algorithm, performs slightly better than its non-reciprocated version, with larger performance gap at later times, \wdyn \text{}, the algorithm with time-varying affinity matrix, outperforms its non-reciprocated equivalent at all time steps.\\

\begin{table}[!htbp]
\caption{\bf{Edge prediction in synthetic networks.} The stronger performance of \dmtrep \text{} in the low-reciprocity regime, $\eta=0.05$, is revealed by looking at the percentage of samples where \dmtrep \text{} has higher AUC than \dmtrepo, on a trial-by-trial case, over 20 trials.}
\begin{adjustbox}{angle=0}
\resizebox{0.99\columnwidth}{!}{%
\begin{tabular}{rrrrr}
\toprule
& \multicolumn{2}{c}{\wdyn}  &
      \multicolumn{2}{c}{\wsta} \\ 
\toprule
T &   \dmtrep &\dmtrepo & \dmtrep & \dmtrepo  \\
\midrule
 1 &             0.0 &              0.0 &            57.0 &             43.0 \\
 2 &            71.0 &             29.0 &            43.0 &             57.0 \\
 3 &            86.0 &             14.0&            38.0 &             62.0  \\
 4 &            67.0 &             33.0 &            43.0 &             57.0 \\
 5 &            71.0 &             29.0 &            52.0 &             48.0\\
 6 &            81.0 &             19.0&            57.0 &             43.0 \\
\bottomrule
\end{tabular}
}
\end{adjustbox}
\label{tab:synth_edge}
\end{table}

Although both variants of the algorithm give better performance than their non-reciprocated version, it could be seen from \cref{fig:AUCsynth}  that  \wdyn \text{}   is more robust in link prediction tasks as $\eta$ increases, and as the planted evolving structure of the affinity matrix changes from assortative to disassortative over time ($T=4,5,6$).  
\\

\subsection{Real world data: reciprocity/AUC }
\label{sec:RD_rec}

To evaluate the capability of our proposed model in retrieving network features, ‌ we apply the model to real world datasets.  In this case, we first apply the inference algorithm to each time snapshot of  the dynamic real dataset and learn the network's latent variables, i.e., $\Theta$. Then, we use these latent variables as the  input for  the generative model, \cref{sec:app}, to generate dynamic synthetic networks similar to the fitted real datasets.  Thus, we can compare dynamic synthetic networks, here $5$ samples,  and the original network.  In this paper, we study the performance of our model in reproducing  reciprocity as a significant  structural parameter of the network. We implement our algorithm on two social and communication datasets, namely,  Email Eu core network \cite{Leskovec2007}  and Statistics Citation Networks \citep{Ji2016} (see \cref{apx:data} for details on data pre-processing).

\subsubsection*{EU email network} 
Email-Eu-core network (EU) is constructed from internal emails exchanged between members of a large European research institution.  At each time step,  there is a directed edge from $i$ to $j$, if $i$ sent an email to $j$. Reciprocity may play a role in that receiving incoming emails may, or not, trigger a response email, similarly to other types of social communication \cite{aoki2016input}. The recorded dataset spans over a period of $803$ days. However, we studied the dynamics of the dataset  by dividing it in both daily and monthly durations. In the first case, we divide the edges in daily intervals (\eud\text{}); then select the snapshots  from $5$ consecutive days, randomly. In the latter case, the intervals are monthly; we select the snapshots from the first recorded year (\eum\text{}). 

\begin{figure}[htb]  
	\includegraphics[width=0.9\linewidth]{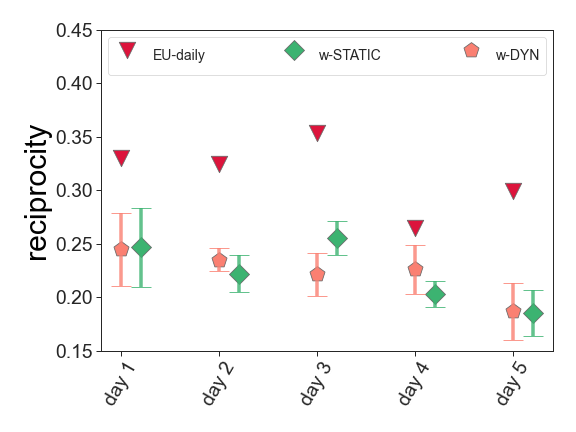}  
	\caption{\textbf{Reproducing the reciprocity of \eud\text{} network.}  Sampled networks were generated based on the inferred parameters fitted to the  \eud\text{} network  \cite{Leskovec2007}. The networks are generated as explained in \cref{sec:app}, with  $N$ and average degree $\langle k \rangle$ as of the real datasets; $K=4$. The markers and  the error bars are the means and standard deviations over $5$ samples of  synthetic networks, respectively.}
	\label{fig:rec_RD_Eu}
\end{figure}

Figure \labelcref{fig:rec_RD_Eu} shows the performance of  \wdyn \text{} and \wsta  \text{} versions of \dmtrep \, in reproducing the reciprocity of the  \eud\text{} network. As expected in email networks, the  reciprocity is high in this case; hence, \wdyn \text{} and \wsta \text{} perform similarly in reproducing reciprocity.  
It is noticeable that  the ability of reproducing reciprocity may change depending on how the network is built. For instance, if we consider the monthly time steps, \eum\text{} network, we observe a different performance, see appendix \labelcref{apx:real_rec}.

Figure  \labelcref{fig:AUCEu} indicates the captured AUCs, measuring performance in link prediction tasks.  The AUC is calculated as described  in \cref{sec:aucsynth}. We can notice the improvement over the time snapshots, and \dmtrep \text{}  tends to perform slightly better. Therefore by having access to the history of the dataset and accounting for reciprocity we can achieve better results in predicting future connections. 

It is worth mentioning that we performed the experiments for different values of the number of communities; however, the results do not show high sensitivity to this parameter. Therefore, we fixed $K=4$ for the EU network, equivalent to the number of departments in the corresponding institute.

\subsubsection*{Statistics Citation dataset} 
The second example of an empirical dataset is the Citation networks for statisticians, which is based on the  research papers published in four of the top journals in statistics from 
$2003$  to the first half of  $2012$.  We construct a network by selecting a sample of the data from $2003$ to $2007$ and dividing it into annual intervals. This way we will have a network of citations over $4$ years, where nodes are authors and an edge from nodes $i$ to $j$ at time step $T$ represents that $i$ cites  $j$'s papers in that year.  In this system, we may expect that reciprocity plays a role in that receiving a citation may trigger a citation back.

\begin{figure}[htb]   
	\includegraphics[width=0.85\linewidth]{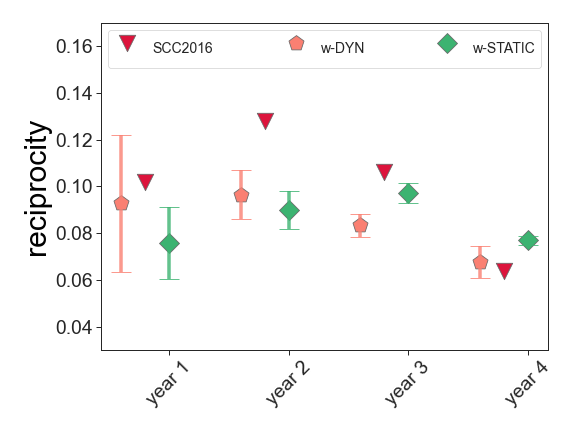}   
	\caption{\textbf{Reproducing the reciprocity of the Statistics citation dataset.}  Sampled networks were generated based on the inferred parameters of the Statistics Citation  dataset \cite{Ji2016}.  The  networks are generated as explained in \cref{sec:app}, with  $N$ and average degree $\langle k \rangle$ as of the real datasets; $K=3$. Markers and bars are the means and standard deviations over $5$ generated synthetic networks, respectively. The network is based on annual citations during four years, from $2010$ to $2013$.}
\label{fig:rec_RD_SCC} 
\end{figure}

Despite the fact that the reciprocity in this dataset is much lower than \eud\text{} dataset, \cref{fig:rec_RD_SCC} shows that we are able to capture it competitively. In addition, although the two versions outperform each others at different time steps, they  still  behave similarly in  reproducing the reciprocity. Moreover, in both empirical datasets, the best performance is obtained for the case that reciprocated edges presented at the same time step were used in the model.

\begin{figure}[htb] 
	\includegraphics[width=0.85\linewidth]{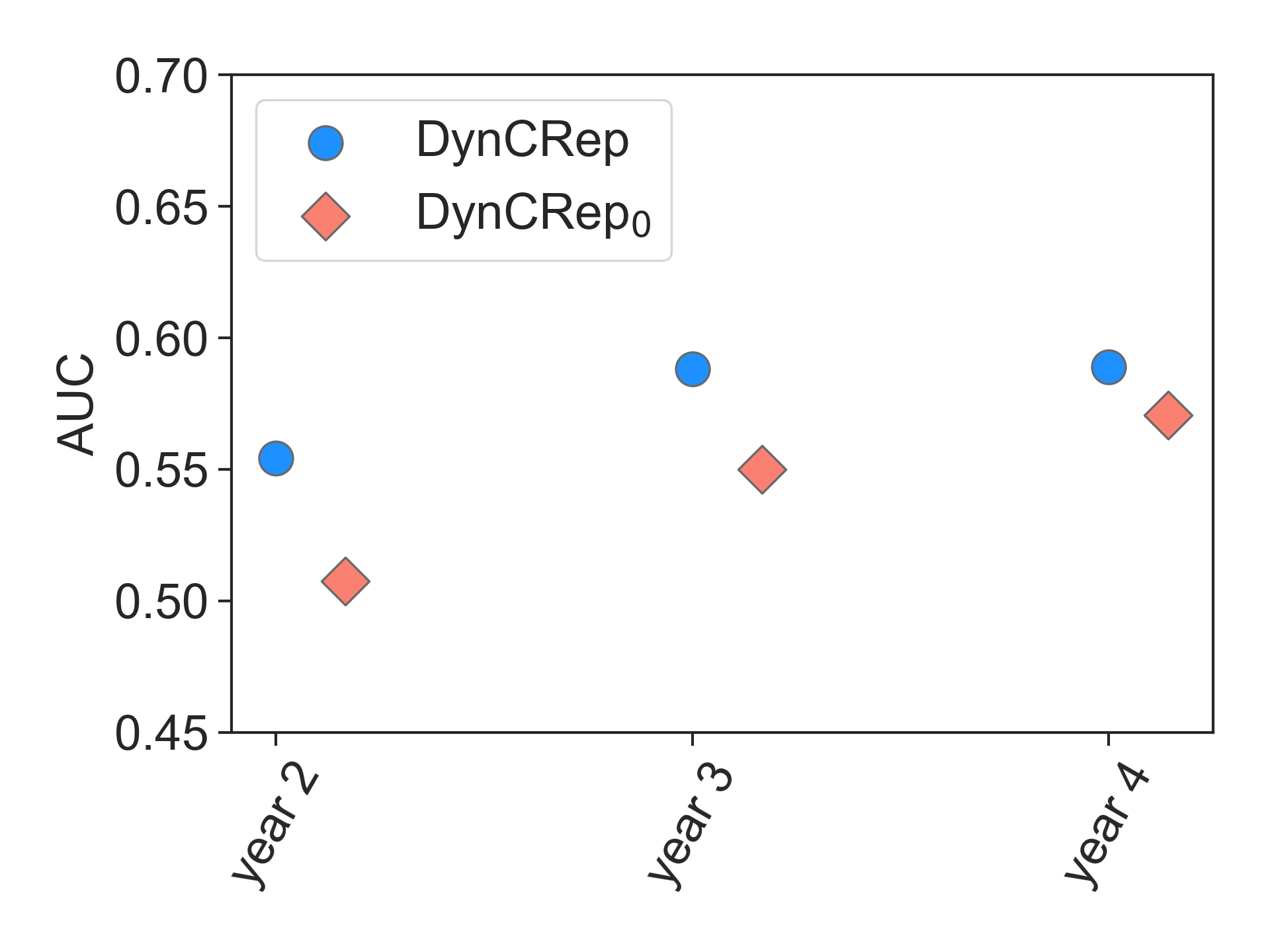} 
	\caption{\textbf{Predicting future evolution in the Statistics citation datasets. }  AUC results for the citation network with $K=3$.  The error bars are too  small to be seen.}
	\label{fig:AUCSCC}
\end{figure}

As it could be seen from \cref{fig:AUCSCC}, AUC values are always higher for \dmtrep \text{}, showing that accounting for reciprocity improves link prediction tasks also for this dataset.
It should be noted that, at each time step $T$ we calculate AUC by having access to the edges up to time $T-1$, then predicting edges at time $T$. Hence, the AUC cannot  be calculated for the first time step. 
In this case we fix $K=3$, the minimum number of communities with the highest performance, i.e., we perform 5-fold cross validation \cite{safdari2020generative} to calculate the value of AUC, then we choose $K$ as the number of communities with the highest value for AUC.
 
\begin{figure}[htb] 
	\includegraphics[width=0.85\linewidth]{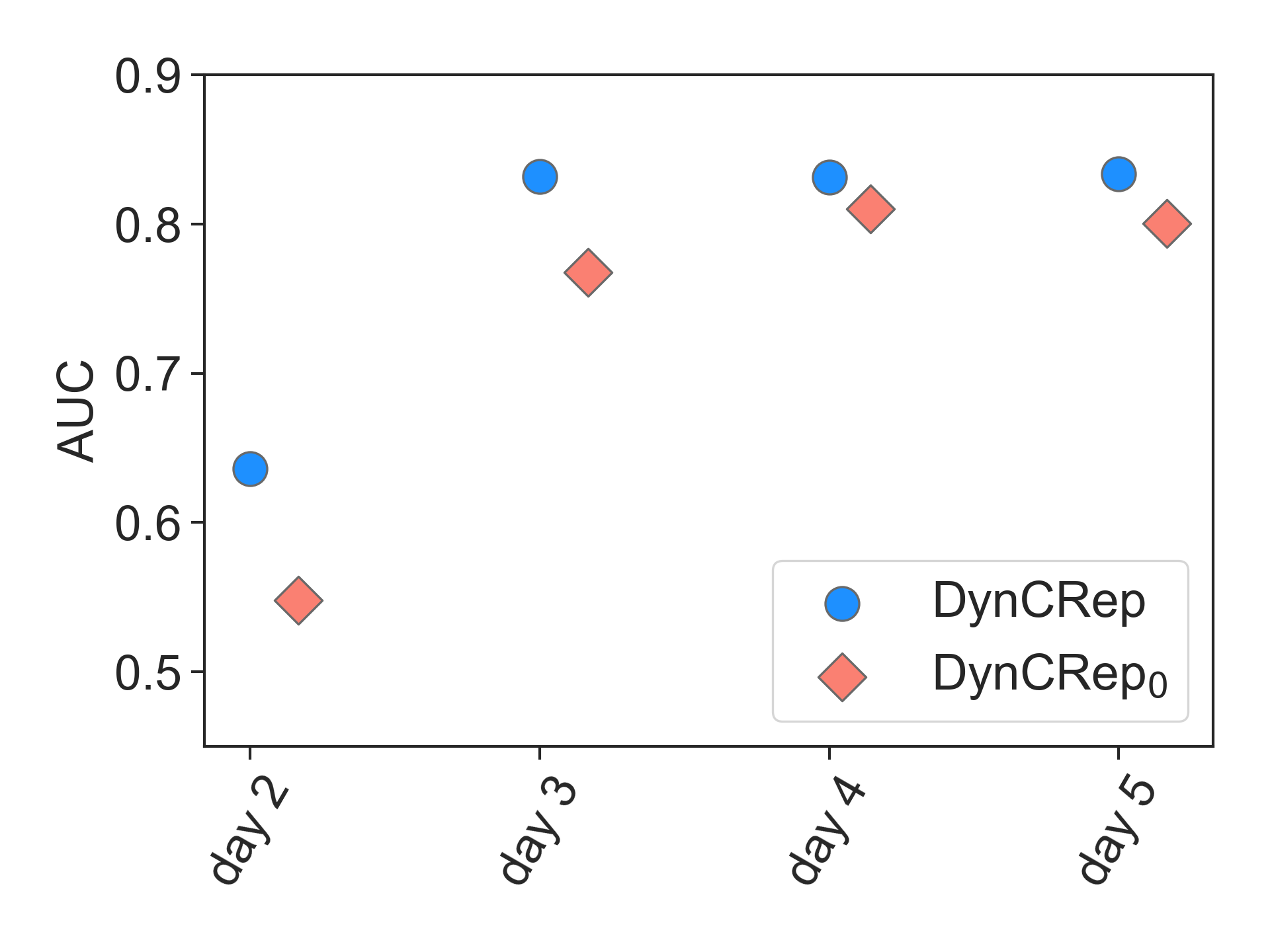}   
	\caption{\textbf{Predicting future evolution in the \eud\text{}  dataset. } AUC results  for \eud\text{} dataset for $5$ consecutive days, selected randomly.  The number of community is fixed to $K=4$. The error bars are smaller than marker size.}
	\label{fig:AUCEu}
\end{figure}

\section{Conclusion}
In this work, we study reciprocity in dynamic networks. In reality, many datasets, e.g., networks of friendship, of gene expression patterns or communication networks, describe interactions that evolve over time, thus making them unsuitable objects of analysis for aggregate methods. In addition, the interactions in these networks might not simply change over time, but their evolution could also be affected by their  past reciprocated   interactions; generally,  such reciprocal interactions have received  little attention as additional drivers of this dynamics.

To remedy this problem, we combine insights from previous works to incorporate reciprocity into a generative model approach with latent community structure.  Specifically ,  we extend the assumptions formulated in \cite{safdari2020generative} to situations where networks change in time. For this, we consider a Markovian transition matrix  which governs  the evolution of the parameters over time snapshots.  
Being a generative model, our approach can be used to build dynamic synthetic networks, with desired reciprocity and community structure. Its algorithmic implementation is based on an efficient EM algorithm, which can be applied to large systems. As we assume a chronological order in observing the reciprocated edges, we can estimate the joint probability distribution as a factorized distribution of time steps. 

We consider two varieties of our model. In one case, community membership vectors remain static over time and only the affinity matrix contains temporal information. In the other case, the affinity matrix is treated as a static parameter, similarly as the community memberships; in both cases, reciprocity parameter and the rate of edge removal are kept static. These two scenarios enable us to  thoroughly  analyze  the model  and its  performance in networks with different interaction patterns.  For instance, in the case of a non-homogeneous  community structure over time, the first  version  would be a more suitable approach, since it could capture the evolving community structures.  

There are a number of directions in which this model could be extended.
To capture more realistic properties of the real world datasets, we can generalize the model to the case of multilayered networks, where nodes can have more than one type of interaction. For instance, in a social network, an individual can have connections based on friendship, as well as her business affiliations.

In addition,  considering a node related reciprocity parameter instead of a global reciprocity parameter could improve the applicability of the model. 
We have focused here on the case where edges change in time, but one can envisage situations where nodes appear and disappear as well. This would also be a natural model extension.
Finally, we considered here reciprocity as main network structural property, but similar investigations can be performed for other properties involving more that one pair of nodes, as triadic closure or transitivity.

\section*{Acknowledgements}
\vspace{-0.1in}
The authors thank the International Max Planck Research School for Intelligent Systems (IMPRS-IS) for supporting Martina Contisciani. 
All the authors were supported by the Cyber Valley Research Fund.  We provide an open source implementation of the code online at \url{https://github.com/hds-safdari/DynCRep}. 


\newcommand{\beginsupplement}{%
        \setcounter{table}{0}
        \renewcommand{\thetable}{S\arabic{table}}%
        \setcounter{figure}{0}
        \renewcommand{\thefigure}{S\arabic{figure}}%
        \setcounter{equation}{0}
        \renewcommand{\theequation}{S\arabic{equation}}
         \setcounter{section}{0}
        \renewcommand{\thesection}{S\arabic{section}}
 }

\clearpage
\newpage 
\bibliographystyle{apsrev4-1} 
\bibliography{bibliography}
    
\clearpage
\beginsupplement 
\begin{widetext}
\section*{{Supporting Information (SI)}}
\section{Master Equation}
\label{appendix:master_eq} 
To solve the master equation in Eq. \ref{eqn:mst1}, we multiply both sides by $z^{k}$ and sum over $k$;  then defining the  generating function $g(z,t)=\sum_{k=0}^{\infty} p^{k}(t) z^{k}$, we get, 
 
\bea\label{eqn:mst-gn}
\f{\partial g(z,t)}{\partial t} &=& \hat{\lambda}_{ij}(t) \, z\,g(z,t)+ \mu  \f{\partial g(z,t)}{\partial z} - \hat{\lambda}_{ij}(t) \,g(z,t)-\mu \,z\, \f{\partial g(z,t)}{\partial z}  \quad \nonumber \\
&& =(z-1)\left[ \hat{\lambda}_{ij}(t)  \,g(z,t)-\mu  \, \f{\partial g(z,t)}{\partial z}  \right]
\eea

By replacing time dependent  $\hat{\lambda}_{ij}(t)$ in Eq. \ref{eqn:mst-gn}, by its expected value, we reach the following relation for $g(z,t)$,

\bea
\label{eqn:gzt1}
g(z,t) =  f\left[(z-1) e^{-\mu t}\right]  e^{\frac{(z-1) \hat{\lambda}_{ij}}{\mu } }, \qquad
\eea
where $f(x)$ is any once-differentiable function of its argument satisfying $f(0) = 1$ \cite{zhang2017random}. 

We can assume that at $t=0$ there was no edges between two specific nodes, hence by setting $t=0$ in Eq. \ref{eqn:gzt1}, 
we find $f(x)= e^{-\f{x  }{\mu} \, \hat{\lambda}_{ij}} $. We assume the mean as $\frac{ \hat{\lambda}_{ij}}{\mu } = \lambda_{ij}+\eta \,A_{ji} $. Therefore, at $t=1$, 

\bea
\label{eqn:gzt2}
g(z,1) &&=   \exp \left\{(z-1) (\lambda_{ij}+\eta \,A_{ji})(1-e^{-\mu })\right\}\nonumber  \\
&& =  \exp \left\{\beta\, (z-1) (\lambda_{ij}+A_{ji})  \right\}  \qquad
\eea

\subsection{Likelihood Maximization}
\label{appendix:EM} 
\bea
L(t, \Theta) &&= \sum _{i,j} \left\{ -\left(\sum _{k,q=1}^K u_{ik} v_{jq} w_{kq}\right)+A_{ij}(0)\log \left[\sum _{k,q=1}^K u_{ik} v_{jq} w_{kq}\right] \right.  \nonumber  \\
&& \left.    +  \sum _{t=1}^T \left[-\beta   \left(\sum _{k,q=1}^K   u_{{ik}} v_{{jq}} w_{{kq}}+\eta  A_{ji}(t-1)\right) +(1- A_{ij}(t-1)) A_{ij}(t)  \log\left[ \beta  \left(\sum _{k,q=1}^K  u_{ik} v_{jq} w_{kq}+\eta\,   A_{ji}(t-1)\right)\right]    \right. \right. \nonumber  \\ \left. \left. \right. \right. \nonumber  \\
&& \left. \left. +A_{ij}(t-1) (1-A_{ij}(t))\log \beta + A_{ij}(t-1) A_{ij}(t) \log[1-\beta ] \right] \right\} 
\eea

To simplify notations and the code implementation, we define:
\bea
\hat{A}_{ij}(t) &=&  \begin{cases} A_{ij}(t) (1-A_{ij}(t-1)) & t>0 \\ A_{ij}(0) & t=0 \end{cases}\\ 
\hat{\beta}(t) &=& \begin{cases} \beta  &t>0 \\ 1 & t=0 \end{cases} \quad.
\eea
With these, the  log-likelihood becomes:
\bea
L(T, \Theta) &&=  \sum_{i,j}\left\{ - \sum_{t=0}^{T} \hat{\beta}(t)  \bup{\sum _{k,q}    u_{{ik}} v_{{jq}} w_{{kq}}+\eta  A_{ji}(t-1)} 
+\sum_{t=0}^{T}\hat{A}_{ij}(t)\log \bup{\sum_{k,q} u_{ik} v_{jq} w_{kq}+\eta \, A_{ji}(t-1)}\right. +  \nonumber  \\
&& \left. +\sum_{t=0}^T \hat{A}_{ij}(t)\log \hat{\beta}(t) +\sum_{t=1} \rup{A_{ij}(t-1) (1-A_{ij}(t))\log \beta + A_{ij}(t-1) A_{ij}(t) \log(1-\beta )} \right\} \quad,
\eea
and we assume that $A_{ji}(-1)=0$.

Because of the summation over variables in the logarithm, in eq. \ref{eq:Lconv}, maximization is not achievable; hence, we apply Jensen's inequality,  $\log \bar{x} \geq \overline{\log x}$, which provides us a lower bound for log-likelihood to maximize. 
\bea
&&\log \left[\lambda^{0}_{ij}+\eta \, A_{ji}(t-1)\right] \geq \rho^{(1)}_{ij}(t) \log \f{\lambda^{0}_{ij}}{\rho^{(1)}_{ij}(t)}+\rho^{(2)}_{ij}(t) \log \f{\eta \, A_{ji}(t-1)}{\rho^{(2)}_{ij}(t)}
\eea
The equality will be satisfied when,
\bea 
&&\rho^{(1)}_{ij}(t)= \frac{\lambda^{0}_{ij}}{\lambda^{0}_{ij}+\eta \,A_{ji}(t-1)} \;,\, \quad \rho^{(2)}_{ij}(t)= \frac{\eta\,A_{ji}(t-1)}{\lambda^{0}_{ij}+\eta\,A_{ji}(t-1)}.\quad    
\eea
As a result, the log-likelihood could be written in terms of these probability distributions, 

\bea
L(T, \Theta) &&  \approx \sum _{i,j}   \left\{- \sum_{t=0}^{T} \hat{\beta}(t)  \bup{\sum _{k,q}   u_{{ik}} v_{{jq}} w_{{kq}}+\eta  A_{ji}(t-1)}  \right. \nonumber  \\
&& \left. +\sum _{t=0}^T \rup{ \hat{A}_{ij}(t)\,\bup{\rho^{(1)}_{ij}(t)\,\log\rup{ \frac{  \sum _{k,q} u_{ik} v_{jq} w_{kq}}{\rho^{(1)}_{ij}(t)}}+\rho^{(2)}_{ij}(t) \log \rup{\frac{\beta\, \eta\,A_{ji}(t)}{\rho^{(2)}_{ij}(t)}}}}  + \sum _{t=0}^T \hat{A}_{ij}(t)\,  \hat{\beta}(t) 
\right.   \nonumber  \\
&& \left. +\sum _{t=1}^T \rup{ A_{ij} (t-1)(1- A_{ij}(t)) \log \beta  + A_{ij} (t-1)A_{ij}(t) \log [1-\beta ] } \right\}. 
\eea

To deal with the summation in the $\log$, we again apply Jensen's Inequality, 
\bea
&&\log \, \beta \sum _{k,q}   u_{ik} v_{jq} w_{kq}  \geq \sum _{k,q}  \phi_{ijkq}  \log \beta \, u_{ik} v_{jq} w_{kq}  - \sum _{k,q}   \phi_{ijkq}  \log  \phi_{ijkq}. 
\eea
The equality  will be established when, 
\be
\phi_{ijkq} =\frac{u_{ik}v_{jq}w_{kq} }{\sum _{k,q}  u_{ik} v_{jq} w_{kq}} \quad.
\ee
It leads to the log-likelihood as a function of the probability distributions, 

\bea
L(T, \Theta) &&=  \sum_{i,j}\left\{ - \sum_{t=0}^{T} \hat{\beta}(t)  \bup{\sum _{k,q}    u_{{ik}} v_{{jq}} w_{{kq}}+\eta  A_{ji}(t-1)} +\sum_{t=0}^{T}\hat{A}_{ij}(t) \bup{ \rho^{(1)}_{ij}(t)\sum_{k,q} \phi_{i,j,k,q} \log u_{ik} v_{jq} w_{kq}  \right. \right.  \nonumber  \\
&& \left.  \left. -\rho^{(1)}_{ij}(t) \sum_{k,q}  \phi_{i,j,k,q}  \log \phi_{i,j,k,q}   +\rho^{(2)}_{ij}(t) \log [\eta \, A_{ji}(t-1)]-\rho^{(1)}_{ij}(t) \log\rho^{(1)}_{ij}(t)-\rho^{(2)}_{ij}(t) \log \rho_2(t)}+\sum_{t=0}^{T} \hat{A}_{ij}(t)\log \hat{\beta}(t) \right.   \nonumber  \\
&& \left. +\sum_{t=1}^{T}\rup{A_{ij}(t-1) (1-A_{ij}(t))\log \beta + A_{ij}(t-1) A_{ij}(t) \log(1-\beta )} \right\} \quad,
\eea

\section{Static parameters (\wsta  \text{})}
\label{APXsec:static}
Assuming static $u,v,w,\eta$ and defining $\hat{A}_{ij}(t) =   A_{ij}(t) (1-A_{ij}(t-1))$ if $t>0$, $\hat{A}_{ij}(0)=A_{ij}(0)$ and $\hat{\beta}(t)=1$ if $t=0$, $\hat{\beta}(t)=\beta$ if $t>0$, we obtain:
\bea
&&u_{ik}= \frac{a-1 + \sum _{j,q,t} \, \rho^{(1)}_{ij}(t)\, \phi _{ijkq}\, \hat{A}_{ij}(t) }{b + \sum _{j,q,t}   v _{jq} \,w_{kq}\,   \hat{\beta}(t) }
= \frac{a-1 + \sum _{j,q,t} \, \rho^{(1)}_{ij}(t)\, \phi _{ijkq}\, \hat{A}_{ij}(t) }{b + \sum _{j,q}   v _{jq} \,w_{kq}\, \bup{1+ \beta\,T}} \label{APXeq:LdifuStatic} \\
&&v_{jq} =\frac{a-1 + \sum _{i,k,t}\, \rho^{(1)}_{ij}(t) \, \phi _{ijkq}\, \hat{A}_{ij}(t) }{b + \sum _{i,k,t} \,   u_{ik}     \,w_{kq} \,   \hat{\beta}(t)}
 =\frac{a-1 + \sum _{i,k,t}\, \rho^{(1)}_{ij}(t) \, \phi _{ijkq}\, \hat{A}_{ij}(t) }{b + \sum _{i,k} \,   u_{ik}     \,w_{kq} \, \bup{1+ \beta\,T}} \label{APXeq:LdifvStatic}\\
&&w_{kq}=\frac{\sum _{i,j,t} \rho^{(1)}_{ij}(t)  \phi _{ijkq} \hat{A}_{ij}(t)}{\sum _{i,j,t}\,  u_{ik}\,v_{jq}\,\hat{\beta}(t) }
=\frac{\sum _{i,j,t}\rho^{(1)}_{ij}(t)  \phi _{ijkq} \hat{A}_{ij}(t)}{\sum _{i,j}\,  u_{ik}\,v_{jq} \, \bup{1+ \beta\,T}} \label{APXeq:LdifwStatic} \\  \nonumber   \\
&&\eta =  \frac{\sum _{i,j,t}\rho^{(2)}_{ij}(t) \hat{A}_{ij}(t)}{\sum _{i,j}\,\sum _{t=1}^{T}\, \hat{\beta}(t) \, A_{ji}(t-1)}=  \frac{\sum _{i,j,t}\rho^{(2)}_{ij}(t) \hat{A}_{ij}(t)}{\sum _{i,j}\,\sum _{t=1}^{T}\, \beta \, A_{ji}(t-1)}, \label{APXeq:LdifeStatic} \\ \nonumber    
\eea

$\beta$ which is also considered a static parameter will be achieved by applying root-finding methods on the following equation, 
\bea
&& \sum _{i,j}  \left\{\sum_{t=1}^T \rup{ -\hat{\beta}(t) \bup{\sum_{k,q} u_{ik} v_{jq}w_{kq} +\eta \, A_{ji}(t-1) }} \right. \nonumber \\
&& \left. + \sum_{t=1}^T\,\hat{A}_{ij} (t)+ \sum_{t=1}^T  \rup{A_{ij}(t-1) (1-A_{ij}(t)) -\f{\beta}{1-\beta} A_{ij}(t-1) A_{ij}(t)}\right\}=0\;,\quad 
\label{APXeq:LdifbetaStatic} 
\eea
with variational distributions:
\bea 
&& \rho^{(1)}_{ij}(t)= \frac{\lambda_{ij}^{0}}{\lambda_{ij}^{0}+\eta \,A_{ji}(t-1)} \;,\quad   \label{APXeqn:rho1static}  \\
&& \rho^{(2)}_{ij}(t)= \frac{\eta\,A_{ji}(t-1)}{\lambda_{ij}^{0}+\eta\,A_{ji}(t-1)}, \quad  \label{APXeqn:rho2static}\\
&&   \phi_{ijkq} =\frac{u_{ik}v_{jq}w_{kq} }{\sum _{k,q}  u_{ik} v_{jq} w_{kq}}\; ,\quad 
\eea
which are time dependent through adjacency matrix.

\section{Only $w$ dynamical (\wdyn \text{}) }
\label{APXsec:w-temp}
Assuming $w(t)$ changing in time while the others remain constant, we have:
\bea 
&&u_{ik}= \frac{a-1 + \sum _{j,q,t}\,\rho _1(t) \, \phi _{ijkq}(t)\, \hat{A}_{ij}(t) }{b + \sum _{j,q}   v _{jq}\, \sum_{t=0}^T  \hat{\beta}(t) \,w_{kq}(t) } \label{eq:LdifuStaticDyn} \\
&&v_{jq} =\frac{a-1 + \sum _{i,k,t}\, \rho _1(t) \, \phi _{ijkq}(t)\, \hat{A}_{ij}(t) }{b + \sum _{i,k}   u_{ik} \, \sum_{t=0}^T  \hat{\beta}(t) \,w_{kq}(t)} \label{eq:LdifvStaticDyn}\\
&&w_{kq}(t)=\frac{\sum _{i,j} \rho _1(t)  \phi _{ijkq}(t) \hat{A}_{ij}(t)}{\sum _{i,j} u_{ik}\,v_{jq} \hat{\beta}(t ) } \label{eq:LdifwDyn}
 \\  \nonumber   \\
&&\eta =  \frac{\sum _{i,j,t}\rho_2(t)  \hat{A}_{ij}(t)}{\sum _{i,j,t=1}  \hat{\beta}(t) \, A_{ji}(t-1)}, \label{eq:LdifeDyn} \\ \nonumber   
 \eea

with the following equation for the log-likelihood:
\bea
L(T, \Theta) &&=  \sum_{i,j}\left\{ -   \sum _{k,q}    u_{{ik}} v_{{jq}}  \bup{\sum_{t=0}^{T} \hat{\beta}(t) w_{{kq}}}+\eta \bup{ \sum_{t=0}^{T} \hat{\beta}(t)  A_{ji}(t-1)} 
+\sum_{t=0}^{T}\hat{A}_{ij}(t)\log \bup{\sum_{k,q} u_{ik} v_{jq} w_{kq}(t)+\eta \, A_{ji}(t-1)}\right. +  \nonumber  \\
&& \left. +\sum_{t=0}^T \hat{A}_{ij}(t)\log \hat{\beta}(t) +\sum_{t=1} \rup{A_{ij}(t-1) (1-A_{ij}(t))\log \beta + A_{ij}(t-1) A_{ij}(t) \log(1-\beta )} \right\} \quad,
\eea
 
$\beta$ as a static parameter could be derived from the following equation by root-finding methods,
\bea
&&\sum _{i,j} \left\{\sum_{t=1}^T \rup{ - \hat{\beta}(t) \bup{\sum_{k,q} u_{ik} v_{jq}w_{kq}(t) +\eta\, A_{ji}(t-1)}} \right. \nonumber \\
&&\left. + \sum_{t=1}^T\,\hat{A}_{ij}(t)+ \sum_{t=1}^T  \rup{A_{ij}(t-1) (1-A_{ij}(t)) -\f{\beta}{1-\beta} A_{ij}(t-1) A_{ij}(t)}\right\}=0\;,\quad 
\eea
with variational distributions:
 \bea 
&& \rho_1(t)= \frac{\lambda_{ij}^{0}(t)}{\lambda_{ij}^{0}(t)+\eta \,A_{ji}(t)} \;,\quad   \label{eqn:rho1staticdyn}  \\
&& \rho_{2}(t)= \frac{\eta\,A_{ji}(t)}{\lambda_{ij}^{0}(t)+\eta\,A_{ji}(t)}, \quad  \label{eqn:rho2staticdyn}\\
&&   \phi_{ijkq}(t) =\frac{u_{ik}v_{jq}w_{kq}(t) }{\sum _{k,q}  u_{ik} v_{jq} w_{kq}(t)} \;,\quad 
\eea
which are time dependent as the result of dependency on adjacency matrix, and $w(t)$.  

\section{Dynamical parameters}
\label{APXsec:dyn}
Assuming $u(t),v(t),w(t)$ changing in time, we need to take only the derivative w.r.t. those individual terms in the loglikelihood. Defining $\hat{A}_{ij}(t) =   A_{ij}(t) (1-A_{ij}(t-1))$ if $t>0$, $\hat{A}_{ij}(0)=A_{ij}(0)$ and $\hat{\beta}(t)=1$ if $t=0$, $\hat{\beta}(t)=\beta$ if $t>0$:
\bea 
&&u_{ik}(t)= \frac{\sum _{j,q}\rho _1(t) \, \phi _{ijkq}(t)\, \hat{A}_{ij}(t)}{\sum _{j,q}  v _{jq}(t)w_{kq}(t)\hat{\beta}(t )} \label{eq:LdifuDyn} \\
&&v_{jq}(t) = \frac{\sum _{i,k}\rho _1(t) \, \phi _{ijkq}(t)\, \hat{A}_{ij}(t) }{\sum _{i,k}  u_{ik}(t)w_{kq}(t) \hat{\beta}(t )} \label{eq:LdifvDyn}\\
&&w_{kq}(t)=\frac{\sum _{i,j} \rho _1(t)  \phi _{ijkq}(t) \hat{A}_{ij}(t)}{\sum _{i,j} u_{ik}(t)\,v_{jq}(t) \hat{\beta}(t ) } \label{eq:LdifwDyn} \\  \nonumber   \\
&&\eta =  \frac{\sum _{i,j,t}\rho_2(t)  \hat{A}_{ij}(t)}{\sum _{i,j,t}  \hat{\beta}(t) \, A_{ji}(t-1)}, \label{eq:LdifeDyn} \\ \nonumber   
 \eea
 with variational distributions:
 \bea 
&& \rho_1(t)= \frac{\lambda^{0}_{ij}(t)}{\lambda^{0}_{ij}(t)+\eta \,A_{ji}(t)} \;,\quad   \label{eqn:rho1dyn}  \\
&& \rho_{2}(t)= \frac{\eta\,A_{ji}(t)}{\lambda^{0}_{ij}(t)+\eta\,A_{ji}(t)}, \quad  \label{eqn:rho2dyn}\\
&&   \phi_{ijkq}(t) =\frac{u_{ik}(t)v_{jq}(t)w_{kq}(t) }{\sum _{k,q}  u_{ik} (t)v_{jq}(t) w_{kq}(t)} .\quad 
\eea

For dynamic $\beta$, we have this equation,
\begin{small}
\bea
\sum _{i,j}\Theta(t-1) \left\{- \beta(t) \bup{\sum_{k,q} u_{ik}(t) v_{jq}(t)w_{kq}(t) +\eta\, A_{ji}(t-1) } + \hat{A}_{ij} (t)+A_{ij}(t-1) (1-A_{ij}(t)) +\f{\beta(t)}{1-\beta(t)} A_{ij}(t-1) A_{ij}(t)\right\}=0\;,\quad   \nonumber  \\
\label{eq:LdifbetaStatic2} 
\eea
\end{small}
where $\Theta_1(t)$, the step function, is $0$ for $t<1$, and $1$ for $t\geq 1$.

\section{Performance in synthetic networks}\label{apx:synth}
\subsection{Link-prediction}
\label{appendixsec:lp}
Figure \ref{appx:fig-AUCsynth_com} compares the ability  of   \dmtrep \text{}  and \mtrep \text{} in the link prediction task. In addition, we tested the model against \dmtrep0 \text{} to evaluate the effect of reciprocity.  The AUC for the \dmtrep \text{} algorithm is obtained as explained in \cref{sec:aucsynth}. For the comparison, we apply \mtrep \text{} algorithm on each time snapshot of the dynamic network, independent from other snapshots. Therefore,  we perform edge prediction using $5$-fold cross-validation. To this end,  at each realization, we divide the dataset, i.e., the entries $A_{ij}(t)$ of the adjacency matrix, into five equal groups selected at random.  We use four of these groups as a training set, to infer the parameters $\Theta$.  We then use the $5$th group as a test set, evaluate the score for each $A_{ij}(t)$ in this set, and calculate the AUC value. By varying which group we use as the test set, we get $5$ trials per realization. The final AUC is the average over these.

As we can see, \wdyn \text{} version of \dmtrep \text{}, by achieving higher values for AUC,  outperforms other approaches in predicting edges. However, when reciprocity effect is higher in dataset ($\eta=0.5$), \mtrep \text{}  could not be suitable model to predict links as it does not deploy the chronological information of the reciprocated edges. 

It is worth mentioning that we also tested the ability of \dmtrep \text{} to recover model parameters $\eta$ and $\beta$. The algorithm shows good results in recovering the value of $\beta$. However, the model underestimates the value of $\eta$ for the initial time steps. But the results improve as we observe more time steps (results not shown).

\begin{figure*}[htb] 
	\includegraphics[width=1\linewidth]{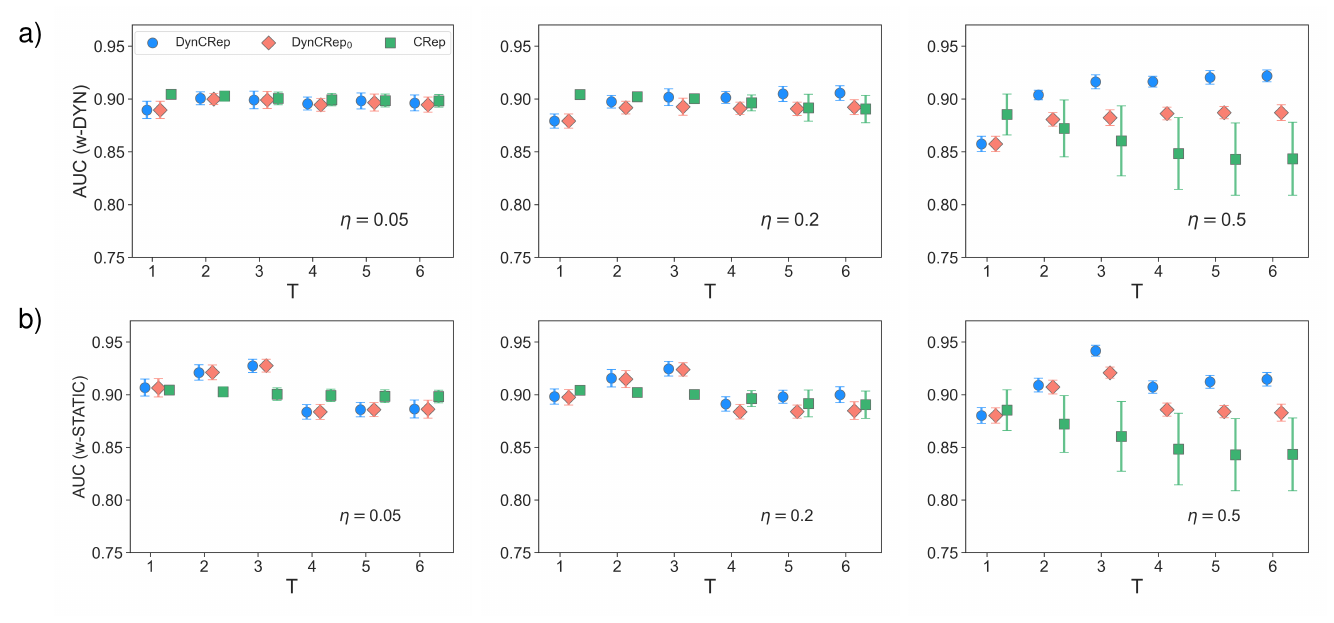}
	\caption{\textbf{Predicting future evolution. } Comparison of the AUC value obtained by \dmtrep \, and the AUC from \mtrep \, applied on each time snapshot. a) \dmtrep \, with time-varying affinity matrix. b) \dmtrep \, with static community membership parameters. In the case of  \mtrep \text{}, AUC was calculated with $5$-fold cross-validation. Networks are generated as explained in \cref{sec:app}, with  $N=500$, average degree $\langle k \rangle=5$, $\beta=0.2$, $K=3$. The three plots are results for $\eta \in \ccup{0.05,0.2,0.5}$. Markers and bars are the means and standard deviations over $20$ network samples, respectively.}
	\label{appx:fig-AUCsynth_com}
\end{figure*}

\clearpage
\section{Performance in real networks}\label{apx:real}
\subsection{Real data: dataset description}
\label{apx:data}
We apply our approach to two different types of  networks: online communication,  and citation networks.
A brief overview of  features of the studied datasets is presented  in the \Cref{tab:apx_data_desc}. 
For the datasets in which the interaction between nodes occur at intervals of varying lengths, we  split  them in time snapshots with equal and suitable time intervals. 
We applied a pre-processing treatment on time snapshots of datasets: i) self-loops  are removed; ii) only nodes that have at least one out-going and one in-coming edge are kept; iii) we used only the giant connected components.
In the case of  citation network (here: SCC2016 ), it requires an additional pre-process of extracting a network of author-author from a network of paper-citation; hence  an edge means that an author cites another author.

\begin{table*}[!h]
\Huge
\begin{center}
\caption{{\bf {Datasets description.}}}
\begin{adjustbox}{angle=0}
\resizebox{1\columnwidth}{!}{%
{\renewcommand{\arraystretch}{1.11}
\begin{tabular}{llllllll}
\toprule
 \textbf{Network} & \textbf{Abbreviation}  &\textbf{Category}  & \textbf{N} & \textbf{E}& \textbf{T}  & \textbf{Ref.}\\
\midrule
Email Eu core network   &EU &Email  Network            & $834$     & $24348$    &$5$        & \cite{Leskovec2007}    \\ 
 Statistics Citation & SCC2016 &Citation Network       & $2654$   & $21568$     &$4$  &    \citep{Ji2016}    \\
\bottomrule
\end{tabular}%
}}
\end{adjustbox}
\label{tab:apx_data_desc}
\end{center}
\end{table*}
\subsection{Reproducing the reciprocity}
\label{apx:real_rec}

\label{apx:data}
\begin{figure}[htb]  
	\includegraphics[width=0.5\linewidth]{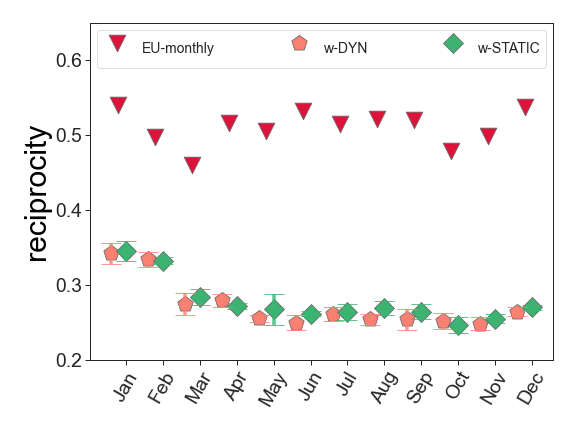}  
	\caption{\textbf{Reproducing the reciprocity of \eum\text{} network.}  Sampled networks were generated based on the inferred parameters fitted on the  \eum\text{} network network \cite{Leskovec2007}. Networks are generated as explained in \cref{sec:app}, with  $N$ and average degree $\langle k \rangle$ as of the real datasets; $K=4$. Markers and bars are the means and standard deviations over $5$ samples of  synthetic networks, respectively.}
	\label{fig:rec_RD_Eu2003}
\end{figure}

\subsection{Inference of parameters $\eta$ and $\beta$.}
\Cref{apx:tbleta}  shows the example of a comparison between the ground truth values of $\eta$ and $\beta$,  applied to generate synthetic datasets,  and the inferred values of the  corresponding parameters by \dmtrep. This shows how inference of the two parameters improves as the number of network snapshots increases, with $\beta$ having values close to the ground truth already for two snapshots. 
\begin{table}[!htbp]
\caption{The inferred values for $\eta$ and $\beta$ by applying the \dmtrep on a synthetic network with: $N=500$, average degree= $20$, $K=3$, $T=6$. }
\begin{adjustbox}{angle=0}
\resizebox{0.60\columnwidth}{!}{%
\begin{tabular}{rrrrr}
\toprule
T & $\eta_0$ &  $\beta_0$  & inferred $\eta$ & inferred $\beta$   \\
\midrule
 1 &            0.5 &             0.2 &            0.01037 &            1.0 \\
 2 &            0.5 &             0.2 &            0.06864&             0.24690 \\
 3 &            0.5 &             0.2 &            0.11148 &            0.24067 \\
 4 &            0.5 &             0.2 &            0.12459&             0.23260 \\
 5 &            0.5 &             0.2 &            0.14086 &            0.22820 \\
 6 &            0.5 &             0.2 &            0.15178 &            0.22496 \\
\bottomrule
\end{tabular}
}
\end{adjustbox} 
\label{apx:tbleta}
\end{table}


\end{widetext}

\end{document}